\documentclass[twocolumn,aps,showpacs,prb,epsf,graphics,psfig]{revtex4}
\usepackage{graphicx}
\usepackage{graphicx}
\usepackage{bbold}
\usepackage{color}
\usepackage[normalem]{ulem}
\usepackage{amssymb}
\usepackage{tikz}
\usepackage[version=3]{mhchem} % \ce{^{.}OH}-radicals
\usepackage{amsmath}

\begin{document}
\title{
Oxygen depletion in FLASH ultra-high-dose-rate radiotherapy: A molecular dynamics simulation
}

%\author{Ramin Abolfath}
%\author{Ramin Abolfath$^{1,2,3,\dagger}$, Santosh KC$^4$, David Carlson$^2$, Alireza G. Senejani$^5$, Alejandro Carabe$^2$, Adri van Duin$^6$, Radhe Mohan, David Grosshans$^{*}$}
\author{Ramin Abolfath$^{1,2,3,a}$, David Grosshans$^{4}$, Radhe Mohan$^{1,b}$}
%, Yusuf Helo$^{4}$, Robert Stewart$^6$, Alejandro Carabe$^2$, Jan Schuemann$^7$, Harald Paganetti $^7$, David Grosshans$^1$, Radhe Mohan$^{1,*}$}
\affiliation{
$^1$ Department of Radiation Physics, University of Texas MD Anderson Cancer Center, Houston, TX, 75031, USA \\
$^2$ Department of Radiation Oncology, University of Pennsylvania, Philadelphia, PA 19104, USA \\
$^3$ Department of Radiation Oncology, New Jersey Urology, West Orange, NJ 07052, USA \\
$^4$ Department of Radiation Oncology, University of Texas MD Anderson Cancer Center, Houston, TX, 75031, USA
%$^1$New Jersey Urology, West Orange, NJ \\
%$^2$Department of Radiation Oncology, University of Pennsylvania, Philadelphia, PA 19104, USA \\
%$^3$Department of Radiation Physics and Oncology, University of Texas MD Anderson Cancer Center, Houston, TX, 75031, USA \\
%$^4$San Diego State University, San Diego \\
%$^5$UNH \\
%$^6$Penn State \\
%$^4$Imanova Centre for Imaging Sciences, Invicro LLC, London, United Kingdom
%$^6$University of Washington, Seattle
}

\date{\today}

%%%%%%%%%%%%%%%%%%%%%%%%%%%%%%%%%%%%%%%%%%%%%%%%%%%%%%%%%%%%%%%%%%%%%%%%%%
\begin{abstract}
{\bf Purpose}:
We present a first-principles molecular dynamics (MD) simulation and expound upon a mechanism of oxygen depletion hypothesis to explain the mitigation of normal tissue injury observed in ultra-high-dose-rate (UHDR) FLASH radiotherapy.

{\bf Methods}: We simulated damage to a segment of DNA (also representing other bio-molecules such as RNA and proteins) in a simulation box filled with H$_2$O and O$_2$ molecules.
Attoseconds physical interactions (ionizations, electronic and vibrational excitations) were simulated by using the Monte Carlo track structure code Geant4-DNA.
Immediately after ionization, {\it ab initio} Car-Parrinello molecular dynamics (CPMD) simulation was used to identify which H$_2$O and O$_2$ molecules surrounding the DNA-molecule were converted into reactive oxygen species (ROS).
Subsequently, the femto- to nano-second reactions of ROS were simulated by using MD with Reactive Force Field (ReaxFF), to illustrate ROS merging into new types of non-reactive oxygen species (NROS) due to strong coupling among ROS.
A coarse-grained model was constructed to describe the relevant collective phenomenon at the macroscopic level on ROS aggregation and formation of NROS agglomerates consistent with the underlying microscopic pathways obtained from MD simulations.

{\bf Results}:
Time-dependent molecular simulations revealed the formation of metastable and transient spaghetti-like complexes among ROS generated at UHDR. At the higher ROS densities produced under UHDR, stranded chains (i.e., NROS) are produced, mediated through attractive electric polarity forces, hydrogen bonds, and magnetic dipole-dipole interactions among hydroxyl (\ce{^{.}OH}) radicals.
NROS tend to be less mobile than cellular biomolecules as opposed to the isolated and sparsely dense ROS generated at conventional dose rates (CDR). We attribute this effect to the suppression of bio-molecular damage induced per particle track. At a given oxygen level, as the dose rate increases, the size and number of NROS chains increase, and correspondingly the population of toxic ROS components decreases. Similarly, at a given high dose rate, as the oxygen level increases, so do the size and number of NROS chains until an optimum level of oxygen is reached.
Beyond that level, the amount of oxygen present may be sufficient to saturate the production of NROS chains, thereby reversing the sparing effects of UHDRs.
%These processes may explain the observed normal tissue sparing effect of UHDRs at physoxic levels ($\approx$ 4-5\%) and the reversal of sparing at high levels of oxygen. They may also explain the lack of FLASH effect for tumors. Generally, tumors have pockets of hypoxia ($<$ 0.4\%) surrounded by regions with higher levels of oxygen. However, sufficiently high tumor dose is prescribed with the aim of eliminating the most-resistant (presumably the hypoxic) fraction. In other words, at UHDRs, oxygenated tumor cells are also rendered hypoxic but killed along with the initially hypoxic cells.  In contrast, initially physoxic normal tissue cells are rendered hypoxic and are spared.

{\bf Conclusions}:
We showed that oxygen depletion, hypothesized to lead to lower normal-tissue toxicity at FLASH dose rates, takes place within femto- to nano-seconds after irradiation. The mechanism is governed by the slow dynamics of chains of ROS complexes (NROS).
Under physoxic ($\approx$ 4-5\% oxygen) conditions (i.e., in normal tissues), NROS are more abundant than in hypoxic conditions (e.g., $<$ 0.3\% in parts of tumors), suggesting that bio-molecular damage would be reduced in an environment with physoxic oxygen levels. Hence irradiation at UHDRs would be more effective for sparing physoxic normal tissues but not tumors containing regions of hypoxia. At much higher levels of oxygen (e.g., $>$ 10-15\%), oxygen depletion by UHDRs may not be sufficient for tissue sparing.
\end{abstract}
%\pacs{87.50.-a, 87.53.-j, 87.55.N-}
\maketitle
%%%%%%%%%%%%%%%%%%%%%%%%%%%%%%%%%%%%%%%%%%%%%%%%%%%%%%%%%%%%%%%%%%%%%%%%%%
\section{Introduction}
\label{intro}
Recent pre-clinical studies have suggested that ultra-high-dose-rate (UHDR) “FLASH” irradiation may be superior to conventional radiation delivery in terms of sparing normal tissues~[\onlinecite{Favaudon2014:STM,Vozenin2018:CCR,Montay-Gruel2018:RO,Montay-Gruel2019:PNAS,Buonanno2019:RO,Vozenin2019:RO,Arash2020:MP}].
In one of those studies, the incidence of lung fibrosis in mice exposed to 4.5 MeV electron beams at UHDR (40 Gy/s or higher) was lower at total doses higher than those used for irradiation at conventional dose rates (CDR).
For example, irradiating mouse lungs to 17 Gy via CDR elicited fibrogenesis at 24 weeks after exposure, in contrast, at FLASH dose rates, only rare fibrotic patches were observed after exposure to 30 Gy~[\onlinecite{Favaudon2014:STM}].
Another study demonstrated that whole-brain irradiation with electrons at dose rates above 100 Gy/s resulted in sparing of brain tissue and cognitive functions in mice~[\onlinecite{Montay-Gruel2018:RO}].
A recent study based on DNA damage to IMR90 human lung fibroblasts from 4.5 MeV proton minibeams (LET = 10 keV/$\mu$m) showed that dose rate had little effect on cell survival, but UHDR irradiation had fewer adverse long-term effects~[\onlinecite{Buonanno2019:RO}].
Specifically, doses of 10-20 Gy delivered at high dose rates (e.g., 1000 Gy/s) led to fewer prematurely senescent cells and lower levels of the pro-inflammatory molecule TGF$\beta$ at 1 month after irradiation compared with dose rates of 0.2 Gy/s~[\onlinecite{Buonanno2019:RO}].

Spitz and colleagues~[\onlinecite{Spitz2019:RO}] used an integrated physico-chemical approach to clarify the relative protection of normal tissue versus tumor tissue after ionizing radiation delivered via FLASH versus CDR.
They proposed a hypothetical mechanism in which FLASH induces a temporary hypoxic environment in normal tissues through oxygen depletion, such that the normal tissues become more radioresistant.
This proposed mechanism described the generation of neutral target radicals after hydrogen abstraction by sparse, spatially distributed hydroxyl radicals (\ce{^{.}OH}).
At UHDR, it was expected that water incorporated by oxygen molecular dissociations happens in highly dense clusters, leading to a non-trivial formation of reactive oxygen complexes and macro-molecules.
Under these conditions, one would expect strong interactions among radiochemical species immediately after their creation, i.e., in the sub-pico-second time scale.
Despite this expectation, the mechanism proposed by Spitz {\em et al.}~[\onlinecite{Spitz2019:RO}]
did not consider interactions among radiochemical species from the chemical and physiological pathways controlled by diffusion rate constants in cells.

Indeed, Koch~[\onlinecite{Koch2019:RO}] took issue with the mechanism proposed by the Spitz group~[\onlinecite{Spitz2019:RO}], contending that additional radical-radical interactions may take place at FLASH UHDR relative to CDR, and that such interactions would have the effect of reducing the biologically effective dose (as opposed to the physical dose).

To examine Koch’s assertion (which is the basis of our hypothesis), we present here the results of our investigation, based on first-principles computations of the molecular dynamics of reactive oxygen species (ROS) immediately (i.e., within pico-seconds [ps]) after their generation.
Our series of hypothetical FLASH-based nanoscale in {\it silico} experiments suggests a microscopic mechanism for oxygen depletion and a process by which, under UHDR conditions, the rate of DNA hydrogen abstraction would slow down simply because ROS, including hydroxyl \ce{^{.}OH}-radicals, substantially screen each other’s effects and tend to form stranded clusters in which hydrogen binding renders them non-reactive (that is, NROS clusters).
As a result, the motion of ROS toward the DNA is counteracted by a dragging force induced electrostatically among ROS that forces them to be localized in space instead of free thermal diffusion of ROS to bio-molecules and DNA.
According to this model, we predict that (1) ROS increase as the radiation dose increases, (2) ROS decrease as the dose rate increases from CDR to UHDRs for a given dose, and (3) there exists an optimal oxygen level that leads to maximum gain in terms of FLASH radiation therapy sparing normal tissues.

In this paper, our simulation steps, including calculation of ionizing and non-ionizing molecular excitations in a particle track structure by using the Geant4-DNA Monte Carlo (MC) tool box~[\onlinecite{Incerti2010:IJMSSC}] followed by molecular dynamics (MD) simulations~[\onlinecite{Abolfath2011:JPC,Abolfath2013:PMB}] are described in Section~\ref{MatMeth}.
Our DNA model and the molecular structure of its surrounding environment, including the distribution of water and oxygen molecules, are described in Section~\ref{DNA}.
Analytical steps in modeling coupled ROS–NROS macroscopic rate equations and our numerical approach to calculating their solutions is provided in Section~\ref{ROS}.
The results of molecular dynamics simulations focusing on ROS aggregation and formation of NROS agglomerates, and the results of numerical solutions of their instantaneous populations induced by a pulse of radiation, are presented in Sections~\ref{Res} and~\ref{PulseDR}.
Discussion and conclusions are presented in Sections~\ref{Diss} and~\ref{Conc}.

\section{Materials and Methods}
\label{MatMeth}
To model radiation track structure, we used Geant4-DNA MC simulations to calculate the spatial distributions of ions in and among simulation boxes and to characterize the dependence of DNA-damage pattern on the beam source, quality, and dose rate.
In the Geant4-DNA MC simulations, particle pencil beams were simulated by irradiating a MC volume, i.e., a series of protons impinging on a cylindrical or cubical water phantom.

The Geant4-DNA package allows the event-by-event simulation of the particle shower produced during the transport of electrons, protons and other radiation modalities in a continuous model of liquid water. In the present version of Geant4-DNA, the static structure of the water molecules is taken from the scattering cross-sections.
However, to study the effect of dose rates on the formation and the evolution of ROS over time, one must incorporate the molecular representation of water, in addition to storage of ionization and molecular excitation coordinates within the volume of MD simulation box, as we did in the present study.

Because of the large number of ionization coordinates and limitations on data storage and analysis of the coordinates, we sampled MD against approximately 100 proton tracks.
We used 100 protons to perform ensemble averaging. Out of these 100 protons we superimposed combinations of a few proton tracks in nm-scale simulation box randomly to study the effect of track-track interactions throughout their chemical products. We did not consider a situation that all 100 protons pass through an MD simulation box simultaneously as this is a highly unlikely configuration.
Note that the aim of this simulation was to demonstrate the creation of NROS owing to ROS coupling via MD. A qualitative Geant4-DNA simulation that is achievable with 100 protons is adequate for such demonstration.

To simulate the dose rate, e.g., at UHDR, the proton tracks were considered to pass through water after a very brief time lapse.
For lower dose rates, we considered larger time lapse among the tracks.
In practice, to simulate the dose rate in nm-size MD volume, where protons travel through MD volume within almost the same interval of time (i.e., attoseconds)
to increase the dose rate (i.e., ionization density), it was convenient to increase the number of incident protons, or equivalently superimposing tracks of protons in the MD volumes.

We further simulated reactive MD by using the Reactive Force Field (ReaxFF) method~[\onlinecite{Abolfath2011:JPC,Abolfath2013:PMB}] implemented in LAMMPS (Large-scale Atomic/Molecular Massively Parallel Simulator), January 2019 version~[\onlinecite{Plimpton1995:JCP,Plimpton1997:SIAM}].
%Details about the MD simulation performed with ReaxFF, including the kind of input data and the "free parameters," are given below.
The ReaxFF method provides the parametric interatomic potential among atoms, i.e., the bond order/bond distance relationship formally introduced by Abell {\em et al.}~[\onlinecite{Abell1985:PRB}], needed to calculate the evolution of molecules over time.
In MD simulations, the trajectories of all atoms in the system are calculated by integrating the equations of motion; hence the forces on the atoms are derived from the ReaxFF potential.
The ReaxFF approach can also accurately describe bond breakage and bond formation, including energies, transition states, reaction pathways, and reactivity trends of both covalent bonds as well as ionic bonds and intermediate interactions.
Simulations on a number of compounds and majority of elements in the periodic table showed that the properties of these compounds and the elements calculated with ReaxFF-MD is in agreement with quantum mechanical (QM) calculations and experiments~[\onlinecite{Yusupov2012:NJP}].

In ReaxFF, the total system energy is the sum of several partial energy terms with regards to valence and torsion angles of molecules that are characteristics of the molecular topology.
Also molecular bond defects and disorders such as lone pairs, under-coordination, over-coordination, conjugation and hydrogen bonding are included in the energy terms.
Moreover, Coulomb and van der Waals energy terms that are part of non-bonded interactions, are taken into account in total energy as  these interactions are calculated between every pair of atoms.
Thus the ReaxFF potential as such is capable of describing covalent bonds and ionic bonds and the entire range of intermediate interactions.
Also the charge distributions based on molecular geometry and intermolecular conductivities are calculated based on electronegativity equalization method~[\onlinecite{Yusupov2012:NJP}] with force field parameters described in~[\onlinecite{Rahaman2011:JPCB}].

Simulation of DNA damage induced by the ionization-track structure was done by selective partitioning of macroscopic MC volumes and mapping them onto a three-dimensional lattice of molecular simulation boxes~[\onlinecite{Abolfath2013:PMB}].
We used cylinders and cubes of various sizes for the MC volumes, depending on the nominal energy of the particle in the beam. The largest geometrical shapes for water phantoms were a cylinder with a 10 cm radius and 20 cm height and a cube measuring 20 cm $\times$ 20 cm $\times$ 20 cm.

To fit a double-stranded DNA structure with sufficient numbers of water and oxygen molecules surrounding the DNA fragment, we set the dimensions of the MD simulation box at 2.6 nm $\times$ 2.6 nm $\times$ 6 nm.
The 6-nm vertical dimension of the simulation domain and the selected length of the section of the DNA molecule were optimized to balance the minimum size of a DNA to fit one double-strand break per DNA length and the largest fragment of DNA, including the surrounding water molecules, that were allowable in the MD simulation.

The 6 nm vertical dimension of the simulation domain and the selected length of the section of the DNA molecule are optimized to balance between the minimum size of a DNA to fit one DSB per DNA-length and the largest fragment of DNA including the surrounding water molecules, allowable in the MD simulation.

The diffusibility of \ce{^{.}OH}-radicals must also be taken into account to estimate a reasonable cut-off for the lateral dimensions of the simulation domain. When the lateral size of the simulation domain increases, the number of water molecules surrounding the DNA increases, leading to unnecessary increases in simulation time. Consequently, we chose the lateral dimension of the simulation domain to be 2.6 nm, to balance the need to minimize computational time (by choosing a simulation domain that is as small as possible) with the need to include sufficient layers of water molecules that are within the diffusion length of the \ce{^{.}OH}-radicals. The optimal value of the lateral dimension of the simulation domain was determined by optimizing the number of water molecules surrounding the DNA to properly incorporate the diffusion of \ce{^{.}OH}-radicals within the ns time scale.

All simulations were performed under periodic boundary conditions, and thus the results reported here presumably are not sensitive to variations in the size of the computational boxes. To test this assumption, we systematically increased the lateral dimensions and observed no significant difference in the end-point results.

The MD simulations used to examine the stability of DNA structure (in the absence of ROS) were done with time steps of 0.25 fs (femto-second) and run for up to 1 ns. The results we present here focus on a range of 100 ps, that is, a period shorter than 1 ns. We chose 0.25 fs as the optimal length of the time step in an attempt to balance the MD run time (very short time steps translate into long run times) with avoidance of artificial instabilities in the DNA double-helix structure (characteristic of very long time-steps).

First-principles calculations based on {\it ab initio} Car-Parrinello MD (CPMD) and density functional theory (DFT)~[\onlinecite{hohengerg1964:PRB,khonsham1965:PRA}] were used to determine the molecular topology and geometric structures of H$_2$O and O$_2$ as a function of their charge defects. In particular, we mapped the separation distance among their constituent atoms to the molecular ionization states induced by the passage of particle tracks. Immediately after ionization, the molecules expand and become unstable if the number of lost electrons exceeds a threshold number owing to electrostatic repulsion among the nuclei and lack of charge neutrality in the ionized state of the molecular structures. Thermal fluctuations, which are incorporated in MD, may assist with such quantum instabilities.

In summary, we started with  the attosecond snapshot of ionization pattern calculated by Geant4-DNA followed by CPMD to determine the initial state of H$_{2}$O and O$_{2}$ molecules, i.e., their molecular topology, geometry, and inter-atomic distances. We then used MD to simulate the time evolution of the entire system. Under quantum and thermal fluctuations, positively charged H$_{2}$O and O$_{2}$ molecules may disintegrate into atomistic hydrogen and oxygen fragments. During the subsequent pico-seconds, the atomic hydrogen and oxygen interact and rebind together through hydrogen bonds to form clusters (NROS) in a series of non-equilibrium (i.e., excited molecular) states.

The rest of this section is devoted to discuss the compatibility between the time-scale of the chemical species induction by passage of a particle in MD computational box and the time-scales used in MD simulations.
In particular we focus on a question whether the chemical species should be gradually embedded in 1 ns MD simulations. %in which case, the uncertainty of the chemical production could be larger than the uncertainty of the MD time steps, such as 0.25 fs, used in this study.

To address this question quantitatively let us calculate the time scale, $dt$, for a particle that passes through a MD simulation box, assuming that the production of chemical species occurs continuously during $dt$ according to the continuous slowing-down approximation (CSDA).
For simplicity, let us consider the particle motion along the largest side of MD box (equal to 6 nm) at the entrance of the beam. Because particles move close to the speed of light at the beam entrance, we find that
$dt = 6 {\rm nm} / (3 \times 10^{17} {\rm nm/s}) = 2 \times 10^{-17} {\rm s} = 20$ as (attosecond).
This time scale is several orders of magnitude shorter than 1 ps to 1 ns to which we simulated DNA-damage using MD. Meaning  that on a ps to ns time scales, all ionizations generated by passage of a single particle can be assumed to occur instantaneously. Notably, the $dt$ of 20 as is also shorter than our MD time step, which is 0.25 fs.

As the particle slows down, the time $dt$ increases. In the case that $dt$ becomes comparable to ns or even ps, the assumption that all ionizations occur simultaneously breaks down.
To examine this situation, let us consider $dt = 1$ ps in which one can find the particles motion at a speed, $v$, several orders of magnitude slower than speed of particles at the beam entrance.
To calculate $v$ we recall the following simple kinematic relation:
%For example, the velocity, $v$, of a proton to pass through 6 nm in 1 ps can be calculated by:
$1 {\rm ps} = 6 {\rm nm} / v$ therefore $v = 6 \times 10^{-9} {\rm m}/10^{-12} {\rm s} = 6 \times 10^3 {\rm m/s}$.
The spectrum of particles moving with such range of velocities corresponds to very low energies.
For protons moving with $v = 6000 m/s$, the kinetic energy, $KE$, can be calculated according to the following non-relativistic relation
$KE = m_p v^2/2 = (1.67 \times 10^{-27} {\rm kg}) (6000 {\rm m/s})^2 / 2 = 30 \times 10^{-21} {\rm J} = 30 \times 10^{-21} \times 6.242 \times 10^{18} {\rm eV} = 0.187 {\rm eV}$.
Note that the conversion factor and constants used in this “back-of the-envelope” calculation are:
$m_p = 1.6726219 \times 10^{-27} {\rm kg}$ and $1 J = 6.242 \times 10^{18} {\rm eV}$.
A kinetic energy of  0.187 eV is very low energy for protons, and thus this scenario (that a proton passes MD box within 1 ps and continuously deposits energy) can be ignored.

A second question may be raised in regards to the simulation of the chemical induction by multi-tracks and compatibility of MD time scales with the multi-tracks time lapse in low dose rates.
However because at UHDR, the proton tracks are considered to pass through water after a very brief time lapse, the assumption that the ionization patterns generated by multi-tracks in a simulation box within a time lapse shorter than 0.25 fs might be reasonable.
For lower dose rates, however, the time lapse of two sequential tracks passing through a single MD simulation box may exceeds the MD time steps. In that case, the chemical induction by multi-tracks must be added to MD sequentially.
This requires developing an interactive Geant4-DNA–MD computational platform. We are indeed in the process of constructing such a platform, which we plan to report in future publications, given that this process is beyond the scope of the present theoretical study aiming to present a hypothesis regarding a macroscopic mechanism for the oxygen depletion at UHDR-FLASH (referring to Eqs. \ref{eq8}-\ref{eq9} below).

To clarify the time steps used in our MD simulations, we reiterate that our computational method contains three-steps: (1) Attoseconds ionization processes; (2) fast disintegration of H$_2$O and O$_2$ molecules, and (3) relatively slow ROS production processes, in addition to their dynamics, ROS-ROS coupling and NROS generation. After scoring ionization events with Geant4-DNA (step \#1), we performed a time-independent ab {\it initio} calculation to determine disintegration of molecules (step \#2) and the scoring chemical
events including forming and breaking bonds using ReaxFF MD that resulted in the generation of ROS and NROS (step \#3).
Thus, the origin of the time for MD simulation with a time step corresponding to 0.25 fs starts immediately after disintegration of H$_2$O and O$_2$ molecules. We used 0.25 fs to divide the ps to ns time frame for simulation of ROS generation, immediately after disintegration of H$_2$O and O$_2$.
Thus, a few hundreds ps production of a specific chemical complex in nature was simulated after few thousands of ReaxFF MD time steps, each with 0.25 fs time interval.

%The core and valence electrons interactions were described within projector-augmented plane-wave (PAW) potentials as implemented in the Vienna \textit{Ab-initio} Simulation Package (VASP)\cite{Heyd2004}. The exchange potential with the generalized gradient approximation of Perdew, Burke, and Ernzerhof (PBE) \cite{Perdew1996}. An energy cutoff of 500 eV is used for the plane-wave basis set in all the calculations. Spin polarization were used in all the calculations. In the PAW potentials used the 2s$^{2}$ 2p$^{4}$, 3s$^{2}$ 3p$^{3}$, 2s$^{2}$ 2p$^{2}$, 1s$^{1}$ and 2s$^{2}$ 2p$^{3}$ electrons were explicitly treated as the valence electrons for O, P, C, H, and N, respectively. First, the  Oxygen molecule in a simulation box of size $15\AA \times 15\AA \times 15\AA$ was optimized. Electrons are gradually removed from the system in order to observe the oxygen bond dissociation. Similary, we have investigated the effect of electron extraction in Guanine and DNA molecules. The charged molecules were relaxed until the Hellman-Feynman forces were less than 0.01 eV/\AA.

The assumption that all ionizations happen in step \#1 may break down in a situation where the average time lapse, $\delta t$, between two-tracks passing through an MD simulation box exceeds the MD time step, $\Delta t = 0.25$ fs.
If $\Delta t \leq \delta t$ the present approach must be extended to a more general model in which a gradual incorporation of ionizations during MD simulation must be performed.
Addressing this issue requires developing an interactive Geant4-DNA–MD computational platform. We are actually in the process of constructing such a platform, which we plan to report in future publications.
%, given that this process is beyond the scope of the present theoretical study.

Thus, the knowledge on the average time-lapse between two tracks that violates the criteria and assumptions in step \#1 is needed to be known.
To this ends we present a simple calculation to estimate an average upper bound for the time-lapse between two tracks that preserves the conditions in step \#1.
Such upper bound in the time-lapse is equivalent to a lower bound in the velocity of particles that are located within few nm to each other.

For simplicity we consider a pencil beam of particles.
Let us consider two moving particles along the longest side of an MD box with a dimension corresponding to $dx = 6$ nm.
Roughly at most a 6 nm separation is required between the two particles if they are restrained to be within a volume of MD-simulation box simultaneously.
Here we show if the speed of these two particles exceed $8 \times 10^{-2}$ times the speed of light, they hit the MD box within 0.25 fs time lapse as $dt =  dx/v =  (6 nm)/v \leq 0.25$ fs.
Hence, $v \geq  (6 nm)/(0.25 fs) =  (6 \times 10^{-9}  m)/(0.25 \times 10^{-15} s) = 24 \times 10^6$ m/s.
Dividing $v$ by speed of light $c$ we obtain $\beta = v/c \geq (24 \times 10^6  m/s)/(3 \times 10^8  m/s)=8 \times 10^{-2}$.
Because the speed of particles at the entrance is close to the speed of light, the above condition can be fulfilled for a significant portion of the tissues at depth if two particles move at the proximity of each other at the lateral plane (perpendicular to the beam central axis). The latter condition can be inferred from the lateral distribution of particles in a beamlet. We note that one should validate this assumption against a MC simulation of the beamlets used in FLASH-therapy using realistic source geometry and particles phase space.

\subsection{The DNA model}
\label{DNA}
\begin{figure}
\begin{center}
\includegraphics[width=1.0\linewidth]{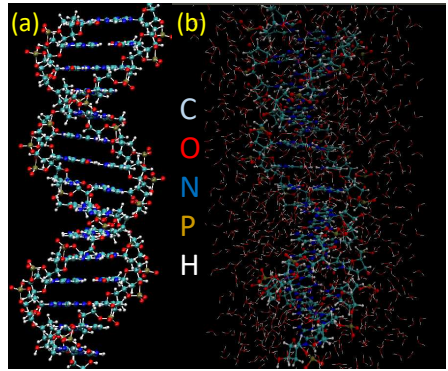}\\ %\vspace{-0.5cm} \\
\noindent
\caption{
DNA (a) in vacuum and (b) in solution. Atom colors are as follows: carbon, cyan;  oxygen,red; nitrogen, blue; phosphorus, gold; and hydrogen, white.
}
\label{fig0}
\end{center}\vspace{-0.5cm}
\end{figure}

The structure of DNA, extracted from a protein database (pdb)~[\onlinecite{Zegar1996:PDB}] and used for our molecular simulations, is shown in Figure~\ref{fig0}a; the same DNA structure surrounded by water and oxygen molecules is shown in Figure~\ref{fig0}b. We filled the DNA environment with water and oxygen molecules corresponding to a water density equal to 1 g/cm$^3$ by using the computational tools available in the Groningen Machine for Chemical Simulations molecular-dynamics package (GROMACS), which is another MD package to simulate proteins, lipids, and nucleic acids~[\onlinecite{Berendsen1996:CPC}].

The excited water and oxygen molecules were simulated to initially generate ROS and NROS. The locations and the type of excitations, i.e., ionization vs. non-ionization excitations were calculated by using Geant4-DNA. Under ionization of water and oxygen molecules, these molecules undergo excited states as they lose electrons. Depending on the molecular ionization states, they subsequently undergo molecular instabilities that lead to disintegration of the molecules into atomic states. Other excited states, with charge neutrality (non-ionization channel), that allow molecular instability and disintegration were taken into account.

We validated the integrity of DNA structure by running ReaxFF-MD for up to 1 ns~[\onlinecite{Abolfath2011:JPC,Abolfath2013:PMB}] which yielded the DNA structure and the equilibration with the solvated molecules. We observed during this time-frame that the DNA kept its overall helical configuration, indicating that ReaxFF retained the overall structural integrity of the DNA over this timeframe and that reactive events observed because of ROS can indeed be associated with the reactivities induced by ionizing radiation exposure.

\subsection{Reactive oxygen species coupling}
\label{ROS}
We next constructed a model for a system of coupled ROS. The first kind of ROS considered in this model includes light species, such as \ce{^{.}OH}-radicals, atomic oxygen, and hydrogen peroxide ($H_2O_2$), as they can diffuse rapidly to DNA and other biomolecules and contribute to processes such as hydrogen abstraction in a very short time. The second kind of species in this model includes relatively higher-mass NROS, such as transient complexes and resonant/metastable clusters (i.e., the OH-transient-clusters) that form via hydrogen bonds immediately after disintegration of O$_2$ and H$_2$O molecules.

$N_1$ denotes the density (number per unit volume) of a set of ROS light species, and $N_2$ denotes the density (number per unit volume) of a set of higher-mass NROS. For an ensemble of ROS that grow and decay, we can construct a series of macroscopic rate equations that describe the time evolution of the size distribution of clusters or complexes of ROS. In a simplistic model, the coupled rate equations can be expressed in the following forms~[\onlinecite{Amar1994:PRB}]:
\begin{eqnarray}
\frac{dN_1}{dt} = G - 2D_f N^2_1 - D_f N_2 N_1,
\label{eq8}
\end{eqnarray}
and
\begin{eqnarray}
\frac{dN_2}{dt} = D_f N^2_1.
\label{eq9}
\end{eqnarray}
In Eq.(\ref{eq8}), which describes variations in the number of ROS of the first kind (i.e., light species), $G$ is a ROS-generating
rate constant, that is proportional to ionizing radiation dose rate $\dot{z}$.
The constant of proportionality, $\mu_1$, depends on the initial density of $O_2$ and $H_20$ molecules, i.e., $G = \mu_1(N_{O_2}, N_{H_2O}) \dot{z}$, and can be used for phenomenologic studies of oxygen effects in tissues, e.g., the hypoxic vs. oxic conditions.

The second term in Eq.(\ref{eq8}) describes the decay in the population of ROS of the first kind due to their recombination with a rate proportional to $N^2_1$, e.g., a recombination process in which a few \ce{^{.}OH}-radicals form OH-transient-clusters.
Note that $N^2_1$ describes a process in which two ROS combine and make one NROS.
$D_f$ in the various terms on the right-hand side of Eqs. (\ref{eq8}) and (\ref{eq9}) resembles the diffusion constant of ROS of the first kind (light species).
The factor of 2 accounts for two ROS of the first kind as they form a cluster of ROS of the second kind.
The third term in Eq.(\ref{eq8}) describes decay in $N_1$ via capturing a ROS of the first kind by clusters of ROS of the second kind, e.g., adsorption of \ce{^{.}OH}-radicals to form metastable OH-transient-clusters.

The time variation in number of ROS of the second kind is described by Eq.(\ref{eq9}), where $N_2$ increases as the new clusters or complexes are formed by coalescence of the ROS of the first kind.
Thus the growth rate of ROS of the second kind is proportional to the probability that two ROS of the first kind meet each other.
For example, the process by which a few \ce{^{.}OH}-radicals interact to form one OH-transient-cluster is given by $D_f N^2_1$ on the right-hand side of Eq.(\ref{eq9}).

For a constant dose rate $\dot{z}$, $G$ is constant.
It is then convenient to transform Eqs.(\ref{eq8}-\ref{eq9}) into dimensionless units, throughout the typical length and time scales of the ROS of the first kind
\begin{eqnarray}
\ell_1 = \left(\frac{D_f}{G}\right)^{1/6},
\label{eq10}
\end{eqnarray}
and
\begin{eqnarray}
t_1 = \frac{1}{(D_f G)^{1/2}}.
\label{eq11}
\end{eqnarray}
Rewriting the rate equations using the dimensionless variables $\tilde{t} \equiv t / t_1$, $\tilde{N}_1 \equiv N_1 \ell^3_1$, and $\tilde{N}_2 \equiv N_2 \ell^3_1$, we obtain
\begin{eqnarray}
\frac{d\tilde{N}_1}{d\tilde{t}} = 1 - 2 \tilde{N}^2_1 - \tilde{N}_2 \tilde{N}_1,
\label{eq12}
\end{eqnarray}
and
\begin{eqnarray}
\frac{d\tilde{N}_2}{d\tilde{t}} = \tilde{N}^2_1.
\label{eq13}
\end{eqnarray}

Analytical solutions of Eqs.(\ref{eq12}-\ref{eq13}) can be obtained in two limits of short and long times.
For sufficiently short times such that $\tilde{t} << 1$, the density of both first and second kinds ROS is small, so that the last two terms in Eq.(\ref{eq12}) can be neglected compared with unity.
This yields $\tilde{N}_1 \propto \tilde{t}$ which can be written in original variables with the physical dimensions
\begin{eqnarray}
N_1 \propto G t.
\label{eq14}
\end{eqnarray}
Insertion of Eq.(\ref{eq14}) into Eq.(\ref{eq13}), we find $\tilde{N}_2 \propto \tilde{t}^3$, which translates to
\begin{eqnarray}
N_2 \propto G^2 D_f t^3.
\label{eq15}
\end{eqnarray}
Note that, according to Eqs.(\ref{eq14}-\ref{eq15}), the dependence of the number of ROS of the first and second kind on $G$ and the ionizing radiation dose rate turns out to be linear and quadratic, respectively, hence $N_2 / N_1 \propto G$. Therefore at low dose rates, $G = \mu_1 \dot{z} << 1$, $N_2 << N_1$, whereas at UHDR, $G = \mu_1 \dot{z} >> 1$, we obtain $N_2 >> N_1$.

Based on Eqs.(\ref{eq14}-\ref{eq15}), over longer time intervals, the population of ROS decreases relative to NROS, which shows that $N_2$ increases much faster than $N_1$ as a function of time. In these limits, because $N_1 < N_2$, we may neglect the second term in favor of the third term in Eq.(\ref{eq12}).
These last terms are, however, relevant over long intervals where the solutions of Eqs.(\ref{eq12}) asymptotically approach the stationary solutions, $d\tilde{N}_1/d\tilde{t} \approx 0$.
Equivalently, $1 - \tilde{N}_1 \tilde{N}_2 \approx 0$, hence $\tilde{N}_1 \tilde{N}_2 \approx 1$, and so $\tilde{N}_1 \approx 1/\tilde{N}_2$.
By insertion of the latter into Eq.(\ref{eq13}), we obtain $\tilde{N}_2 \propto (3\tilde{t})^{1/3}$, leading to the expression:
\begin{eqnarray}
N_2 \propto G^{2/3} D^{-1/3}_f t^{1/3}.
\label{eq16}
\end{eqnarray}

For the pulse-shape dose rates of the types used for UHDR, there is no analytical solution to Eqs.~(\ref{eq8}-\ref{eq9}), thus requiring a numerical integration over these equations.

\subsection{Remarks on dose and dose rate at the microscopic vs. macroscopic scales}
This subsection is devoted to elucidating the calculation of dose and dose rate at the microscopic scales using MD simulation boxes.
It is important to distinguish the differences between physical quantities that are well defined at the macroscopic level and their equivalence at the microscopic level.
The “temperature of an object” is a classical textbook example of such quantities that are measurable ``only” at the macroscopic level.
At the microscopic / atomistic level, the temperature loses its applicability.
Instead, the kinetic energy of the constituent atoms is the well-defined physical quantity.
An ensemble average over the kinetic energy of the atoms in a finite-size system (in a relevant coarse-grained model) must be performed to scale up and calculate the temperature of that object.

Similar steps must be performed for dose and dose rate in our problem, as the dose and dose rate are dosimetrically measurable quantities at the macroscopic level. In an atomistic model, however, we can only talk about the energy transfer to a single atom/molecule.

With regards to the calculation of ionization energy in a type of simulation that we have performed, and conversion to energy per unit mass at nano-meter scales in unit of Gy, it is not possible to make a meaningful comparison with experiments without knowing the macroscopic details of the beams.
Here we present a detailed example of why this is the case:
Let us consider a single computational box with nm-scale dimension (as described in our MD simulations, e.g., a box with dimension of 2.5 nm $\times$ 2.5 nm $\times$ 6 nm).
If we fit approximately 1200 H$_2$O molecules in that box, we construct a finite size system of water molecules with a bulk mass density corresponding to 1000 kg/m$^3$.
Let us consider that, in this box, we score only one ionization event to an H$_2$O molecule.
For a ballpark estimate, let us further consider 13 eV (21 $\times$ 10$^{-19}$ J) for ionization energy per water molecule.
%Subtracting the kinetic energy of the released electron, exited from the MD simulation box, yields the energy deposition in the respective volume of MD simulation box.
For conversion to the deposited dose, one must divide the energy deposition by the mass inside the volume, a subtlety at the atomistic level.
%Which mass of molecules should be accounted for the denominator to convert the energy deposition to Gy?
Even if all 1200 water molecules are accounted for a bulk mass enclosed in the volume, a large number in Gy will result (note that in this example, we only consider a single ionization event to an H$_2$O molecule).
More explicitly, the deposited dose is given by $E_{\rm tr} / {\rm mass} =  21 \times 10^{-19} {\rm J} / (1200 \times 30.2 \times 10^{-27} {\rm kg}) = 57,947$ Gy, neglecting the kinetic energy transferred to the released ($\delta$-ray) electrons, exiting from the MD computational box.
Here $E_{\rm tr}$ denotes energy transfer to a single water molecule and $30.2 \times 10^{-27}$ kg approximately accounts for approximately the mass of a single H$_2$O molecule.

It is important to recognize that the result of this back-of-the-envelop calculation does not imply that the deposited dose at the macroscopic level would be 57,947 Gy because of strong heterogeneity in the distribution of $E_{\rm tr}$ among nm-scale computational boxes.
%Indeed, a knowledge on the macroscopic volume under irradiation is necessary in mapping the beam to atomistic ionization tracks, in addition to the number of simulation boxes that occupy that macroscopic volume.
%The calculation of dose for such a system is computationally challenging, because dividing a cm-scale volume to nm-scale MD simulation boxes end up a number that exceeds 10$^{19}$ or 10$^{20}$.
Among 10$^{19}$ to 10$^{21}$ nm-scale computational boxes that fill a cm-size volume, a significant number are hit with zero tracks.
These volumes contribute negligibly to $E_{\rm tr}$.
On average, however, the deposited dose calculated by sum of $E_{\rm tr}$ over all MD boxes divided by their total mass, would converge to the experimental values.

Note that the purpose of MD models by design is to simulate chemical processes at the atomistic level hence they are not convenient for calculation of macroscopic quantities such as the dose and dose rate.
An extensive computational work must be performed to scale up meaningfully the sum of energy transfers from atomistic level to dose at the macroscopic level.
In the present work, however, the dose and dose rate were incorporated in a coarse-grained model, as given by Eqs. (\ref{eq8}) and (\ref{eq9}).
These equations, however, were constructed based on underlying microscopic phenomenon inferred from MD simulations.

\section{Results}
\label{Res}
Track structures simulated by Geant4-DNA MC, followed by {\it ab initio} CPMD and ReaxFF-MD simulation reveal induction of chemical species that are less efficient in causing DNA-damage at FLASH-UHDRs because of the environment induced at such dose rates~[\onlinecite{Abolfath2011:JPC,Abolfath2013:PMB}].

\begin{figure}
\begin{center}
\includegraphics[width=1.0\linewidth]{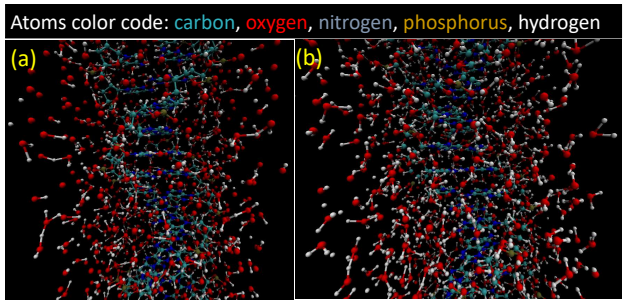}\\ %\vspace{-0.5cm} \\
\noindent
\caption{
Initial structure of DNA in solution (a) with oxygen (i.e., aerobic conditions) and (b) without oxygen (i.e., hypoxic conditions). As shown in panel (a), oxygen atoms or molecules are present, whereas this not the case in panel (b). Atoms colors are as follows: carbon, cyan; oxygen, red; nitrogen, blue; phosphorus, gold; and hydrogen, white.
}
\label{fig1_1}
\end{center}\vspace{-0.5cm}
\end{figure}

\begin{figure}
\begin{center}
\includegraphics[width=1.0\linewidth]{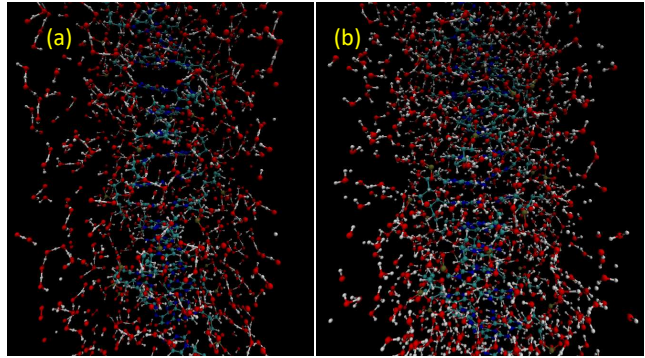}\\ %\vspace{-0.5cm} \\
\noindent
\caption{
Final structure of DNA in solution after 1 ps molecular dynamic (MD) simulation at room temperature, starting from the initial structures shown in FIG.~\ref{fig1_1} under (a) aerobic and (b) hypoxic conditions. The DNA double- helix molecule is surrounded by reactive oxygen species (ROS). Populations of chain-like OH-transient-clusters (i.e., metastable/resonant molecules) in the aerobic condition are more numerous than in the hypoxic condition (compare left and right corners of image (a) with image (b)). See also FIG.~\ref{fig2}. Atom colors are as follows: carbon, cyan; oxygen, red; nitrogen, blue; phosphorus, gold; and hydrogen, white.
}
\label{fig1}
\end{center}\vspace{-0.5cm}
\end{figure}

\begin{figure}
\begin{center}
\includegraphics[width=1.0\linewidth]{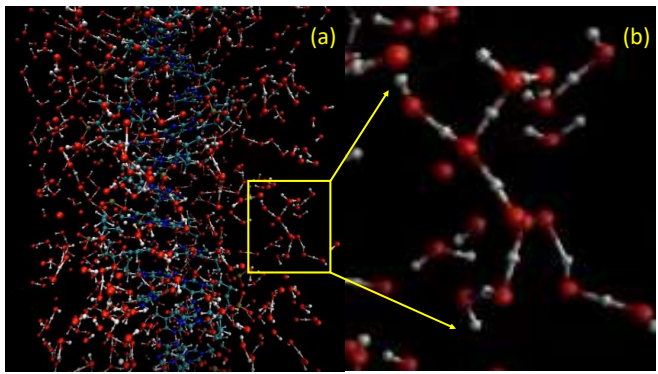}\\ %\vspace{-0.5cm} \\
\noindent
\caption{
Panel (a) at left shows the transition state of DNA and its environment at 1 ps after irradiation. Molecular dynamics simulation was done with ReaxFF MD for up to 1 ns in 0.1- fs time steps at the high ionization density relevant to FLASH ultra-high-dose-rate (UHDR) radiation. Panel (b) at right shows chain-like OH complexes formed from interactions of H$_2$O and O$_2$ molecules at UHDR, representing non-reactive oxygen species (NROS). These complexes form spontaneously around the DNA molecule at UHDR, suggesting that the generation of \ce{^{.}OH}-species can be considered a global phenomenon, i.e., is not limited to accidental local formation and sparse distribution of NROS at conventional dose rates.
}
\label{fig2}
\end{center}\vspace{-0.5cm}
\end{figure}

\begin{figure}
\begin{center}
\includegraphics[width=1.0\linewidth]{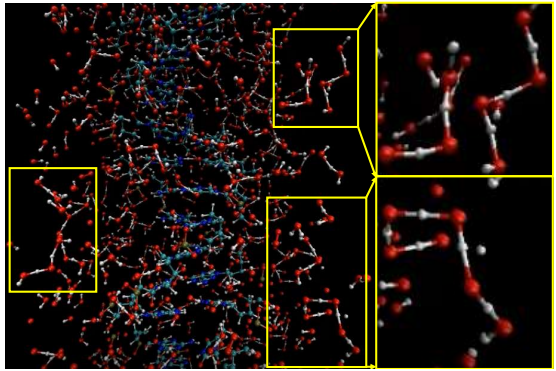}\\ %\vspace{-0.5cm} \\
\noindent
\caption{
Molecular dynamics simulation at higher oxygen concentrations but below the FLASH-ultra-high-dose-rate (UHDR) reversal point. Chain-like OH complexes are formed from interactions of H$_2$O and O$_2$ molecules at UHDR. As oxygenation increases, the number of chain-like OH complexes increases and the ability of FLASH irradiation to damage DNA decreases. If a typical tumor is more hypoxic than normal tissues, then FLASH would act to spare normal tissues and to damage tumors in ways comparable to damage from conventional dose rate radiation.
}
\label{fig5}
\end{center}\vspace{-0.5cm}
\end{figure}

Figure~\ref{fig1_1} shows typical initial structure of the molecules (DNA, H$_2$O and O$_2$) used in our simulations where the number of H$_2$O and O$_2$ molecules can be tuned to mimic various oxygen levels in tumors and normal tissues, i.e., hypoxic, physoxic and oxyic conditions.
Figures~\ref{fig1}-\ref{fig5} show the corresponding final states, calculated from the time-evolution of the initial states as given in Figure~\ref{fig1_1}.
The likelihood that ROS generated in the vicinity of each other are less mobile and less reactive compounds is illustrated in Figures~\ref{fig1}-\ref{fig5}.
To differentiate the mobilities of isolated ROS vs. lumps of ROS (NROS) the real time evolution of the molecular configurations were uploaded in series of videos to the first author's facebook page, available in a link to the following public domain~[\onlinecite{Ramin_fb}].

Therefore, for a given dose, fewer ROS are available at UHDR. Moreover, the simulations shown in Figures~\ref{fig1}-\ref{fig5} reveal that both oxygen and OH depletion take place simultaneously. Because of the compactness of ROS generated at FLASH, many ROS combine to produce less reactive and more stable chemical complexes. This is the mechanism that we refer to as oxygen depletion in this study.
We interpret these simulations as indicating that less damage to DNA takes place under FLASH than under CDR, resulting in less normal-tissue toxicity. In conditions where regions of hypoxia exist, such as in tumors, and where the dose is prescribed to kill the most resistant cells, this mechanism (oxygen depletion) is less productive, and so tumors show less sensitivity at UHDR than do normal tissues.

%To better visualize the processes described above, videos corresponding to Figures~\ref{fig1}-\ref{fig5} are included as supplementary material and have been uploaded to the first author’s Facebook page~[\onlinecite{Ramin_fb}].

The other important aspects of these simulations is the role of oxygen in the nearby DNA and other bio-molecules. Based on our quantum chemistry calculation, the ionization of O$_2$ with the loss of three or more electrons, forces the molecule to break down into two oxygen atoms. Because atomic oxygen is highly reactive to bio-molecules, when O diffuses and grabs one H from DNA to become an \ce{^{.}OH}-radical, a second hydrogen abstraction is expected to take place immediately after the first one. However, because, after the first transition, the DNA locally transitions to a higher energy state, it becomes more resistant to losing the second hydrogen to form a water molecule. Therefore, the \ce{^{.}OH}-radical formed diffuses a longer distance and reacts with other segments of the same bio-molecule or causes hydrogen abstraction from a different bio-molecule. However, under FLASH UHDR conditions, these \ce{^{.}OH}-radicals may be trapped by other ROS and removed from their path toward second hydrogen abstractions. This is a second mechanism that may reduce the toxicity to normal tissues.

Molecular simulations that include systematic variations in oxygen content (Figures \ref{fig1_1}-\ref{fig2}) revealed an intriguing process by which the population of ROS decreases with increase in oxygen content below a transition point, beyond which the effect of oxygen at FLASH-UHDR is reversed. Within this range of oxygen content, this process suggests that DNA damage at a given FLASH UHDR is lower in an oxyic environment than in a hypoxic environment.

To illustrate the oxygen effect, consider a transition from $H_2O$ molecules to chains of OH-transient-clusters.
In the absence of oxygen, the ratio of H to O atoms is 2 to 1.
This ratio limits populations of OH-transient-clusters, as the need is greater for an excess of hydrogen atoms than for oxygen atoms.
With an increase in oxygen content, however, this ratio decreases.
Hence the number of oxygen atoms available to form chains of OH-transient-clusters increases.
Therefore, in presence of adequate oxygen molecules, the probability of forming OH-transient-clusters increases, which results in lower toxicity to DNA and other bio-molecules.

Heuristically, one may infer an optimal value for oxygen content such that for two molecules of H$_2$O, one molecule of O$_2$ maximizes the chance of formation of a chain with four \ce{^{.}OH}-radicals. Up to this optimum value of oxygen, one may expect to observe more protection to normal tissues from FLASH. Above the optimum value the sparing effects of FLASH would be lost.

%\begin{figure}
%\begin{center}
%\includegraphics[width=1.0\linewidth]{Fig4.pdf}\\ %\vspace{-0.5cm} \\
%\noindent
%\caption{
%Final state of DNA and its environment at 1 ps.
%}
%\label{fig3}
%\end{center}\vspace{-0.5cm}
%\end{figure}
%
%\begin{figure}
%\begin{center}
%\includegraphics[width=1.0\linewidth]{Fig5.pdf}\\ %\vspace{-0.5cm} \\
%\noindent
%\caption{
%Another view at 1 ps, generated by rotating Fig. \ref{fig3} around DNA axis.
%}
%\label{fig4}
%\end{center}\vspace{-0.5cm}
%\end{figure}

\begin{figure}
\begin{center}
\includegraphics[width=1.0\linewidth]{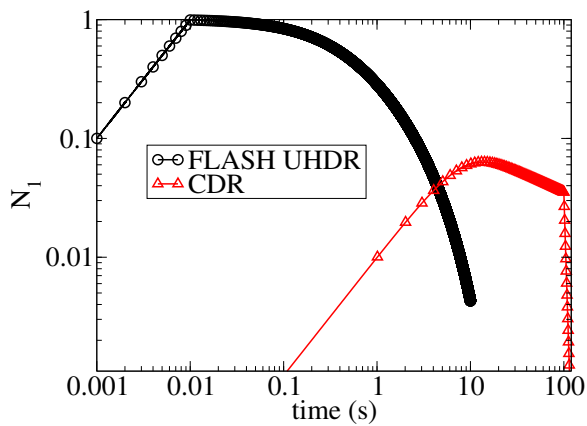}\\ %\vspace{-0.5cm} \\
\noindent
\caption{
$N_1$, the population of reactive oxygen species (ROS) as a function of time and dose rate. At the initial time of ultra-high-dose-rate (UHDR) irradiation, the ROS population predominates. However, the greater the increase in $N_1$ by ionizing radiation, the more quickly the ROS population decreases after radiation. Thus, at conventional dose rates (CDRs), the population of ROS is higher during the typical period of DNA repair, and thus tissue toxicity is greater at CDR than at UHDR. The initial time ($t=0$) for both FLASH-UHDR and CDR are the same. Because of log-log scales in the plots, the initial time was omitted from the graph.
}
\label{fig6}
\end{center}\vspace{-0.5cm}
\end{figure}

\begin{figure}
\begin{center}
\includegraphics[width=1.0\linewidth]{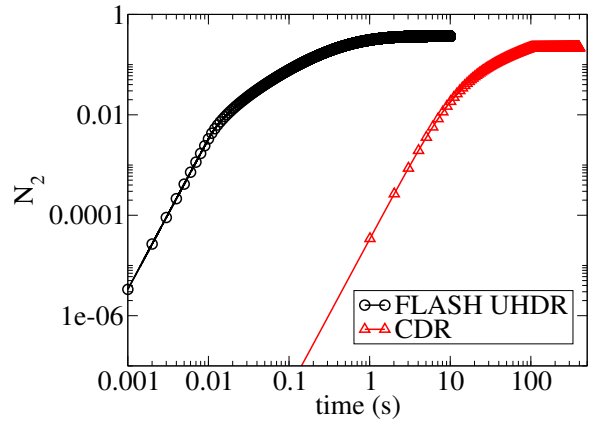}\\ %\vspace{-0.5cm} \\
\noindent
\caption{
$N_2$, the population of non-reactive oxygen species (NROS) as a function of time and dose rate. At longer times, $N_2$ at UHDR is approximately twice that at CDR for the specific pulse used in this calculation. The higher the population of NROS at UHDRs, the lower the toxicity to normal tissues. As shown in FIG.~\ref{fig6}, the initial time ($t=0$) for both FLASH-UHDR and CDR are the same. Because of log-log scales in the plots, the initial time was omitted from the graph.
}
\label{fig7}
\end{center}\vspace{-0.5cm}
\end{figure}

\subsection{Pulse dose rates}
\label{PulseDR}
The results for pulse-shape dose rates of the types used for FLASH-UHDR in Eqs.~(\ref{eq8}-\ref{eq9}) are illustrated in Figures~(\ref{fig6} and \ref{fig7}).

As an example, to distinguish the effects of UHDR (relative to CDR) on $N_1$ and $N_2$, we consider two dose rates in the form of square pulses with amplitudes $G_1=100$ cm$^{-3} s^{-1}$ and $G_2=0.01$ cm$^{-3} s^{-1}$, and width $t=0.01$ s and $t=100$ s.
$G_1$ and $G_2$ are zero outside of these time domains and represent FLASH-UHDR ($G_1$) and CDR ($G_2$).
Our choice for pulse amplitudes and widths is based on the same deposited dose, such that $\int_{0}^{\infty} dt G_1(t) = \int_{0}^{\infty} dt G_2(t) = 1$.
(The numerical values used for $G_1$ and $G_2$ were chosen only to illustrate the method.)

The dependence of ROS populations (denoted by $N_1$) and NROS populations (denoted by $N_2$) on dose rates, are shown in Figs.~(\ref{fig6}) and (\ref{fig7}).

According to these results, at FLASH-UHDR the population of ROS increases sharply over a very short period. However, the greater the increase in $N_1$, the faster the decrease, because of the conversion of $N_1$ to $N_2$. In other words, the amounts of ROS rapidly diminish, to be overtaken by amounts of NROS.

\section{Discussion}
\label{Diss}
Our simulations at UHDR, performed based on a microscopic model, illustrate the events and processes by which oxygen species (including \ce{^{.}OH} radicals, toxic oxygen, H$_2$O$_2$) strongly interact to form chains of oxygen–hydrogen complexes in resonant-like or metastable molecular states that are connected through a series of hydrogen bonds; this result is consistent with Koch’s qualitative description~[\onlinecite{Koch2019:RO}]. These molecules have restricted diffusion capacity and are therefore less likely to interact with and damage DNA. For a single chemical reaction, e.g., a DNA hydrogen abstraction, to take place, a single ROS must diffuse toward the DNA in the presence of collective dragging forces induced by other ROS. Under such circumstances, ROS essentially move in a more “viscous medium,” and are less reactive to DNA. In other words, the contribution of ROS to DNA damage is lower at FLASH dose rates than at CDR.

In the hypothesis proposed by Spitz {\em et al.}~[\onlinecite{Spitz2019:RO}], oxygen depletion is associated with the formation of organic peroxyl radicals (ROO\ce{^{.}}) in a process that seems rather slower than another mechanism in which oxygen grabs or scavenges a hydrogen atom or $e_{aq}$ to prevent the recombination of \ce{^{.}OH} radicals. Using the same terminology as that introduced by Spitz {\em et al.}~[\onlinecite{Spitz2019:RO}], we termed the latter process ``oxygen depletion." Thus we infer that the process of oxygen depletion must be very fast relative to other reactions for this scenario to be dominant.

To investigate the effect of dose rates on population of ROS quantitatively, we considered hydrogen–oxygen complexes in two distinct classes, categorizing light and free complexes such as \ce{^{.}OH} radicals, H$_2$O$_2$, singlet oxygens, superoxide (O$^{2-}$), and alpha-oxygen as ROS in one class and categorizing the other complexes, such as  OH-transient-clusters, which are NROS in a different class. In our coarse-grained (macroscopic) model, we showed that the population of ROS scales linearly with the dose rate, whereas the population of NROS scales quadratically. This model correctly predicted the dominance of the population of NROS at UHDR and the results are consistent with the predictions from our microscopic model.

After the resonance stage that occurs within a brief period, single ROS are released by the disintegration of NROS. Thus one might reasonably expect that these ROS would carry on and cause damage to DNA. However, such a process is not effective because of the lower diffusion of ROS that had already been dragged into the transient clusters. The localization of atomic elements suggests that it is more likely that hydrogen and oxygen atoms recombine into their global stable states and form H$_2$O and O$_2$ before they reach the DNA and cause damage. Thus, the overall damage to DNA is expected to be lower at FLASH dose rates than at CDR.

Our findings based on the dependence of ROS and NROS populations calculated for the two limits of FLASH-UHDR and CDR are consistent with our observations from our MD simulations. Accordingly, at FLASH-UHDR, the population of ROS increases sharply over a very short period. However, the higher the ROS population, the faster it declines because of the conversion of ROS to NROS, leading to rapid diminishment of ROS and supercedence of NROS. At CDR, in contrast, the ROS population is higher during the typical interval when enzymatic DNA repair processes are engaged. Hence, tissue toxicity is likely to be greater at CDR than at UHDR.

Such transitions between the dominance of one kind of ROS (light species) and the other (the more complex and non-reactive species) as function of dose rate are consistent with empirical observations of the effects of FLASH. Both models presented in this study, the MD simulations and the rate equations, predict that the second kind of ROS (NROS) is dominant at UHDR, and thus the bio-chemical reactivity of the byproducts generated at such high dose rates is theoretically lower than at CDR, where the first kind of ROS are dominant.

Montay-Gruel {\em et al.}~[\onlinecite{Montay-Gruel2019:PNAS}] recently reported an interesting observation associated with UHDR FLASH radiation therapy. In measuring H$_2$O$_2$, they detected lower levels at UHDR than at CDR; they further found that the difference between H$_2$O$_2$ levels at UHDR and CDR was considerably greater at higher doses compared with lower doses (10-20 Gy). This group also reported that doubling the oxygen concentration in the brain (by carbogen breathing during irradiation) reversed the neurocognitive benefits of FLASH radiation therapy.

Our predictive model presented in this study captures the overall features of these experimental results~[\onlinecite{Montay-Gruel2019:PNAS}], that (1) ROS increases with an increase in the radiation dose, (2) ROS decreases with an increase in dose rate from CDR to UHDRs at a given dose, and (3) there exists an optimal oxygen level for maximum sparing of normal tissues with FLASH. Given the observed trends in oxygen effects and qualitative interpretation of our predictive model, we believe that increasing the oxygen concentration and the corresponding probability of toxic oxygen formation (including an increase in population of ROS) would lead to ROS-related normal-tissue damage predominating at the oxygen concentrations used in Montay-Gruel’s experiments. This in turn, led to the reversal of the neurocognitive benefits of FLASH radiation therapy.

We remark that, the processes described in this work may explain the observed normal tissue sparing effect of UHDRs at physoxic levels ($\approx$ 4-5\%) and the reversal of sparing at high levels of oxygen. They may also explain the lack of FLASH effect for tumors. Generally, tumors have pockets of hypoxia ($<$ 0.4\%) surrounded by regions with higher levels of oxygen. However, sufficiently high tumor dose is prescribed with the aim of eliminating the most-resistant (presumably the hypoxic) fraction. In other words, at UHDRs, oxygenated tumor cells are also rendered hypoxic but killed along with the initially hypoxic cells.  In contrast, initially physoxic normal tissue cells are rendered hypoxic and are spared.

\section{Conclusion}
\label{Conc}
We present here a qualitative and comparative mechanism to explain the dependence of ROS and NROS populations between the two dose- rate limits of FLASH-UHDR and CDR, independent of the source of ionization. Our focus was to sketch the rates of generation of ROS and their conversion to NROS for two extremely different dose rates. The results of these calculations help to form our hypotheses and set the foundation for future experiments currently being planned. Results from those experiments will form the basis of quantitative estimates of damage to cells and tissues at FLASH-UHDR. We plan to present our quantitative results on DNA damage as a function of different radiation doses, dose rates, energies, and sources of radiation in future publications.

{\bf Acknowledgement:}
The authors gratefully acknowledge useful discussions with Drs. David Carlson, Alejandro Carabe-Fernandez, Alejandro Bertolet Reina, Yujie Chi, Stephen McMahon, Santosh Khatrichhetri (KC), Adri van Duin and Christine Wogan. The work at the University of Texas MD Anderson Cancer Center was supported by the NIH / NCI under Grant No. U19 CA021239.

\noindent{\bf Authors contributions:}
RA and RM wrote the main manuscript. RA prepared the figures and performed mathematical derivations and computational steps. DG and RM proposed the scientific problem and co-supervised the project.

\noindent{\bf Competing financial interest:}
The authors declare no competing financial interests.

\noindent{\bf Corresponding Authors:}\\
$^a$ ramin1.abolfath@gmail.com / Ramin.Abolfath@pennmedicine.upenn.edu  \\
%$^*$ rmohan@mdanderson.org
$^b$ rmohan@mdanderson.org

\end{document}